\documentstyle[preprint,aps,eqsecnum]{revtex} 
\def \p{\par}
\def \en{\end{eqnarray}}
\def \bg{\begin{eqnarray}}
\def \enm{\end{mathletters}}
\def \bgm{\begin{mathletters}}
\def \MO{M_1\theta_{51}}
\def \MT{M_2\theta_{52}}
\def \nd{\noindent}
\def \inf{\infty}

\def \c#1{{\cal#1}}
\def \m#1{{\bf#1}}
\def \ovr{\over}
\def \inf{\infty}
\def \r{\rho}

\def \rh{\hat r}

\def \D{\Delta}

\def \OOT{{1\over 2}}
\def \Orc2{{1\over c^2}}

\def \pri{^{\prime}}
\def \Ph{\hat P}
\def \th{\theta }

\def \Pp{P \cdot p}

\def \prx{\approx}
\def \dt{\cdot}
\def \d{\partial}

\def \gm{\gamma}
\def \a{\alpha}
\def \b{\beta}
\def \e{\epsilon}
\def \xp{x_\perp}
\def \rv{\bbox{r}}

\def \s{\sigma}

\def \Dlt{\Delta}
\def \Oc2{O(1/c^2)}
\def \k{\kappa}
\def \l{\lambda}

\def\uline{\underline}
\begin{document}
\preprint{ORNL-CTP-95-07 hep-ph/9603402} 
\title 
{Singularity-Free Breit Equation 
from Constraint Two-Body Dirac Equations}
\author{
Horace W. Crater}
\address{
The University of Tennessee Space Institute, Tullahoma,
Tennessee, 37388 
}
\author{
Chun Wa Wong}
\address{
Department of Physics and Astronomy, University of California, Los Angeles, CA
90095-1547
}
\author{
Cheuk-Yin  Wong}
\address{
Oak Ridge National Laboratory, Oak Ridge, TN 37831-6373
}
\maketitle
\vspace{2cm}
\begin{abstract}
\noindent
We examine the relation between two approaches to the quantum
relativistic two-body problem: (1) the Breit equation, and (2) the
two-body Dirac equations derived from constraint dynamics. In
applications to quantum electrodynamics, the former equation becomes
pathological if certain interaction terms are not treated as
perturbations. The difficulty comes from singularities which appear at
finite separations $r$ in the reduced set of coupled equations for
attractive potentials even when the potentials themselves are not
singular there. They are known to give rise to unphysical bound states
and resonances. In contrast, the two-body Dirac equations of
constraint dynamics do not have these pathologies in many
nonperturbative treatments. To understand these marked differences we
first express these contraint equations, which have an ``external
potential'' form similar to coupled one-body Dirac equations, in a
hyperbolic form. These coupled equations are then re-cast into two
equivalent equations: (1) a covariant Breit-like equation with
potentials that are exponential functions of certain ``generator''
functions, and (2) a covariant orthogonality constraint on the
relative momentum.  This reduction enables us to show in a transparent
way that finite-$r$ singularities do not appear as long as the
the exponential structure is not tampered with and the exponential generators 
of the interaction are themselves nonsingular for finite $r$. 
These  Dirac or Breit equations,
free of the structural singularities which plague the usual Breit
equation, can then be used safely under all circumstances, encompassing
numerous applications in the fields of particle, nuclear, and atomic 
physics which involve highly relativistic and strong binding 
configurations.

\end{abstract}

\narrowtext
\vfill\eject
\newpage
\section{ Introduction}
\setcounter{section}{1}

In contrast to the accepted description
\begin{eqnarray}
[\gm\dt p+m +V]\psi=0
\end{eqnarray}
of the relativistic quantum mechanics of a single spin-one-half particle
moving in an external potential $V$ given by Dirac, a number
of different approaches have been used for 
two interacting spin-one-half particles.  A traditional
one based on the Breit equation [1], also known as the Kemmer-Fermi-Yang
equation [2],
\begin{eqnarray}
[\bbox{\a}_1\dt\bbox{p}_1+\b_1 m_1+\bbox{\a}_2\dt\bbox{p}_2
+\b_2 m_2+V(\bbox{r}_{12})]\psi=E\psi,
\end{eqnarray}
contains a sum of single-particle Hamiltonians and an interaction term
between them. ($E$ is the total energy in an arbitrary frame.)
Although the Breit equation is not manifestly covariant, it has
provided good perturbative descriptions of the positronium and muonium
energy levels.  However, it is well known that those parts of the
Breit interaction
\begin{eqnarray}
V(\rv)={-\a\ovr r}(1-\OOT\bbox{\a}_1\dt\bbox{\a}_2-\OOT\bbox{\a}_1\dt
\hat {\bbox{r}}\bbox{\a}_2\dt\hat {\bbox{r}}),
\end{eqnarray}
in the Breit equation beyond the Coulomb term should not be treated
nonperturbatively, but must be handled only perturbatively.  In other
words, a consistent treatment of the Breit equation in powers of $\a$
will generate unwanted terms not present in quantum electrodynamics
[Refs. 3,4].  

A more serious difficulty of the Breit equation is that when the
interactions are treated nonperturbatively, structural pole singularities
could appear at finite $r$ even when the interactions themselves are
singularity-free there [5,8].
Two of us have found recently that these pole singularities occur
under certain conditions depending on well-defined algebraic relations
among the different potentials that could appear \cite{won93a,won93b}.
They lead to unphysical states or unphysical resonances and therefore
must be strictly avoided \cite{chi82,ko68}.

The primary purpose of this paper is to show how these pole singularities can 
be avoided from the beginning so that the Breit equation can be used without 
difficulty in diverse applications in particle, nuclear, and atomic physics 
involving highly relativistic motions and strong binding potentials.  This is 
accomplished by relating this older approach to one that has been developed 
much more recently.

Dirac's constraint Hamiltonian dynamics [9] provides a framework for
an approach proposed by Crater and Van Alstine [10,11] that differs
notably from that of the Breit equation.  It gives two-body coupled
Dirac equations, each of which has an ``external potential'' form
similar to the one-body Dirac equation in that for four-vector and
scalar interactions one has 
\bgm \bg \c S_1\psi\equiv\gamma_{51}
[\gamma _1 \cdot (p_1 - \tilde A_1) + m_1 +\tilde S_1] \psi = 0, \en
\bg \c S_2\psi\equiv\gamma_{52} [\gamma _2 \cdot (p_2 - \tilde A_2) +
m_2 +\tilde S_2] \psi = 0,  \en \enm
($\gamma_{5i},\ i=1,2$ are the $\gamma_5$ matrices for the 
constituent particles). 
Unlike the Breit approach, these
equations are manifestly covariant and have interactions introduced by
minimal substitution. 

They have common solutions if the operators
$\c S_1, \c S_2$ commute. This situation, called strong compatibility 
[9], can be achieved by the proper choice of the operators $\tilde A_i$
and $\tilde S_i$ [10,11].  The commutator cannot, however,
be made to vanish for more general types of interactions such as 
pseudoscalar or pseudovector. Nevertheless, under certain 
circumstances the commutator can be reduced to 
a combination of $\c S_1$ and $\c S_2$ themselves. The equations are then 
said to be weakly compatible [9], because 
this will also ensure that solutions 
of $\c S_1\psi=0$ in the more general cases
could be solutions of $\c S_2\psi=0$ as well. These compatibility
properties are important because they guarantee the existence of common
solutions to (i.e. the  consistency of) the constraint equations before 
they are actually solved. 

Although the constraint two-body (CTB) Dirac equations have been used
far less frequently than the Breit equation, they have important
advantages over the latter. In  describing electromagnetic bound states 
[11,12]
they yield nonperturbative and perturbative results in agreement with each other.
That is, the exact (or numerical) solution produces a spectrum that agrees 
through order $\a^4$ with that given by perturbative treatment of the Darwin, 
spin-orbit, spin-spin, and tensor  terms obtained from the Pauli reduction.
In particular, total c.m. energies $w$  for the 
$e^+e^-$ system in the $^1J_J$ states 
is found to satisfy a
Sommerfeld formula [11,12]
\begin{eqnarray}
w&=&m\sqrt{2+2 \biggl / \bigg(1+{\a^2\ovr 
n+\sqrt{(J+{1\over 2} )^2-\alpha^2} -J-\OOT)^2}\bigg)}\nonumber\\
&=& 2m-{m\a^2\ovr 4n^2}-{m\a^4\ovr 2n^3(2J+1)}+{11\ovr 64}{m\a^4\ovr n^4}
+O(\a^6).
\end{eqnarray}
They agree through order $\a^4$ with 
those of the perturbative solution of the same equation, and also 
with those of standard approaches to QED. A recent paper has numerically 
extended this 
agreement at least to the $n=1,2,3$ levels for all allowable $J$ and 
unequal masses \cite{cra93}. 

In this paper, we are concerned with another advantage
of the CTB Dirac equations, namely that no unphysical states and
resonances of the type discussed in [6,7] have ever appeared in past
applications. We are able to show here that this is in fact true for a
general interaction and that this is a consequence of the exponential
structure of the interactions appearing in them. This result is obtained
by first reducing the CTB Dirac equations to a Breit equation and an
equation describing an orthogonality constraint. The equivalent Breit
equation is then shown to be singularity-free, provided that the
exponential interaction structure is not destroyed by inadvertent
approximations and that the operators appearing in the exponent are
themselves free of finite $r$ singularities.

    The exponential structure that tames the unphysical singularities
turns out to be a consequence of a relativistic ``third law'' describing the
full recoil effects between the two interacting particles. We carefully
trace, in the formulation of the CTB Dirac equations, how this structure
arises from the need to make these equations at least weakly compatible.
Compared to the {\it laissez-aller} approach of the Breit equation for
which any interaction seems possible, the restriction of the interaction
structure needed in the constraint approach represents a conceptual
advance in the problem. For this reason, we take pain to elucidate its
conceptual foundation as we present elements of the CTB Dirac equations
needed for our demonstration that they are singularity-free. 

This paper is organized as follows.  We start in Sec. II with a brief overview 
of the derivation of the constraint two-body Dirac equations for scalar 
interactions both to define the 
notation used and to describe the main concepts involved. 
One of the most important properties is the compatibility of the 
two constraints. We remind the reader in Sec. III how to introduce 
general interactions into them that preserve this property.  In Sec.IV 
we derive a covariant version of the Breit equation from the 
constraint equations, introducing the concept of exponential generators 
for a wide range of covariant interactions. In Sec. V we 
clarify the structure of this covariant Breit equation by decomposing it
into a matrix form involving singlet and triplet components of 
the matrix wave function followed by reduction to radial form by the use 
of a vector spherical harmonic decomposition. This reveals clearly how 
the constraint approach avoids the structural pole singularities that have
plagued the original Breit equation since its introduction over 65 years ago. 
In Sec. VI we show by contrast, how the 
pole singularities arises for each of the nonzero 
generators if one uses the original Breit interaction. Section VII contains 
brief concluding remarks.
\section{ 
Two-Body Dirac Equations of Constraint Dynamics}

Following Todorov [13], we shall use the following
dynamical and kinematical
variables for the constraint
description of the relativistic two-body problem:

\hskip 1cm {i.)} relative position,  $\ x_1 - x_2$

\hskip 1cm {ii.)} relative momentum, $\ p =  (\epsilon_2 p_1 -
\epsilon_1 p_2)/w \,, $

\hskip 1cm {iii.)} total c.m. energy, $\ w = \sqrt {-P^2} \,,$

\hskip 1cm {iv.)} total momentum, $\ P = p_1 + p_2 \,,$

\noindent
and ~~\hskip 0.5cm {v.)} constituent on-shell c.m. energies,
\begin{eqnarray}
\epsilon_1 = {w^2 + m_1^2 - m_2^2 \over 2w}, 
~~~~\epsilon_2 = {w^2
+ m^2_2 - m_1^2 \over 2w}.
\end{eqnarray}
\nd
In terms of these variables, we have
\begin{eqnarray}
p_1=\e_1\Ph+p,~~~~p_2=\e_2\Ph-p,
\end{eqnarray}
 where $\Ph=P/w$.

We start from the (compatible) Dirac equations for two free particles
\bgm
\bg
\c S_{10}\psi = (\theta_1 \cdot p_1+m_1 \theta_{51})\psi=0\,,
\en
\bg
\ S_{20}\psi = (\theta_2 \cdot p_2 +m_2 \theta_{52})\psi=0 \,,
\en
\enm
where $\psi$ is just the product of the two single-particle Dirac wave
functions. These equations can be written as
\bgm
\bg
\c S_{10}\psi = (\theta_1 \cdot p + \epsilon_1 \theta_1 \cdot \hat P +
m_1 \theta_{51})\psi=0 \,,
\en
\bg
\ S_{20}\psi = (-\theta_2 \cdot p + \epsilon_2 \theta_2 \cdot \hat P +
m_2 \theta_{52})\psi=0,
\en
\enm
when expressed in terms of the Todorov variables.
The ``theta" matrices
\begin{eqnarray}
\theta^\mu_i \equiv i\sqrt {1 \over 2} \gamma_{5i} \gamma^\mu_i,
 \  \mu=0,1,2,3 \ ,i=1,2
\end{eqnarray}
\begin{eqnarray}
\theta_{5i} \equiv i\sqrt {1 \over 2} \gamma_{5i} 
\end{eqnarray}
satisfy the fundamental anticommutation relations
\begin{eqnarray}
[\theta^\mu_i, \theta^\nu_i]_+ = -  g^{\mu \nu},
\end{eqnarray}
\begin{eqnarray}
[\theta_{5i}, \theta^\mu_i]_+ = 0,
\end{eqnarray}
\begin{eqnarray}
[\theta_{5i}, \theta_{5i}]_+ = -1.
\end{eqnarray}
[Projected ``theta" matrices then satisfy
\begin{eqnarray}
[\th_i\dt \Ph,\th_i\dt\Ph]_+=1, 
\end{eqnarray}
\begin{eqnarray}
[\th_i\dt \Ph,\th_{i\perp}^\mu]_+=0, 
\end{eqnarray}
where $\th_{i\perp}^\mu=\th_{i\nu}(\eta^{\mu\nu}+\Ph^\mu\Ph^\nu)$].
They are modified Dirac matrices \cite{misce1} which ensure that the
Dirac operators $\c S_{10}$ and $\c S_{20}$ are the 
square root operators 
of the corresponding mass-shell operators $-\OOT(p_1^2+m_1^2)$ and
$-\OOT(p_2^2+m_2^2).$  

Using the Todorov variables 
and the above brackets, the difference
\begin{eqnarray}
(\c S_{10}^2-\c S_{20}^2)\psi=0=\OOT(p_1^2+m_1^2-p_2^2-m_2^2)\psi
\end{eqnarray}
leads to an equation
\bg
\Pp\psi=\OOT[w(\e_1-\e_2)-(m_1^2-m_2^2)]\psi=0.
\en
The physical significance of the orthogonality  of these two momenta is to
put a constraint on the relative momentum ( eliminating the
relative energy in the c.m. frame).

We will also use covariant (c.m. projected) versions of the
Dirac $\a$ and $\b$ matrices here defined by
\begin{eqnarray}
\b_i=-\gm_i\dt\Ph=2\th_{5i}\th_i\dt\Ph, 
\end{eqnarray}
\begin{eqnarray}
\a_i^{\mu}=2\th_{i\perp}^{\mu}\th_i\dt\Ph, 
\end{eqnarray}
and 
\begin{eqnarray}
\sigma_i^{\mu}=\gm_{5i}\a_i^{\mu}= 2\sqrt{2}i\th_{5i} \, \th_i\dt\Ph
\, \th_{\perp i},\ ~~i=1,2.
\end{eqnarray}
These covariant Dirac matrices take on the simple form 
$\alpha_i^\mu=(0,\bbox{\alpha}_i)$ and
$\s_i^\mu=(0,\bbox{\sigma}_i)$ in the center-of-mass system for
which $\hat P=(1,\bbox{0})$.

If we now introduce scalar interactions between these particles by naively 
making the minimal substitutions 
\begin{eqnarray}
m_i\to M_i=m_i+S_i,\ i=1,2
\end{eqnarray}
(as  done in the  one-body  equation),
the resulting Dirac equations
\bgm
\bg
\c S_1\psi = (\theta_1 \cdot p + \epsilon_1 \theta_1 \cdot \hat P + M_1
\theta_{51})\psi=0, 
\en
\bg
\c S_2\psi = (- \theta_2 \cdot p + \epsilon_2 \theta_2 \cdot \hat P +
M_2 \theta_{52})\psi=0 
\en
\enm
\setcounter{equation}{18} 
will not be compatible because
\begin{eqnarray}
[\c S_1, \c S_2]_- \psi= [\th_1\dt p,\MT]+[\MO,-\th_2\dt p]-
=- i (\partial M_1 \cdot \theta_1 \theta_{52} +
\partial M_2 \cdot \theta_2 \theta_{51} )\psi \not = 0,
\end{eqnarray}
where $\d$ is the  four-gradient.

In the  earlier work \cite{cra82,cra87}, compatibility is reinstated
by generalizing 
the naive $\c S_1$ and $\c S_2$ operators with the help of supersymmetry 
arguments. The procedure contains four major steps:

\par
{
 a) Find the supersymmetries of the pseudoclassical limit of an ordinary free 
one-body Dirac equation.
}
\par
{
b)  Introduce interactions of a single Dirac particle with
external potentials that preserve these supersymmetries.
For scalar interactions, this requires the coordinate replacement
\begin{eqnarray}
x^\mu\to \tilde x^\mu\equiv x^\mu+i{\th^\mu\th_5\ovr m+ S(\tilde
x)}.
\end{eqnarray}
(Note that the Grassmann variables satisfy $\th^2=0$. As a result this self 
referent or  recursive relation
has a terminating Taylor expansion).
}
\par
{
c) Maintain the one-body supersymmetries for each spinning particle
through the replacement 
\begin{eqnarray}
(x_1-x_2)\to (\tilde x_1-\tilde x_2)
\end{eqnarray}
in the relativistic potentials $S_i$.
}
 
These steps  lead to the 
pseudoclassical constraints (the weak equality sign $\prx$ means these equations
are constraints imposed on the dynamical variables)
\bgm
\bg
\c S_1= (\theta_1 \cdot p + \epsilon_1 \theta_1 \cdot \hat P + M_1
\theta_{51} - i \d M_2/M_1\cdot \theta_2
\theta_{52} \theta_{51})\prx 0,  
\en
\bg
\c S_2 = (- \theta_2 \cdot p + \epsilon_2 \theta_2 \cdot \hat P +
M_2 \theta_{52} + i \d M_1/M_2 \cdot \theta_1
\theta_{52} \theta_{51})\prx0.
\en
\enm
They are strongly compatible
under the following two conditions:

\hskip 1.1cm {i.)} the mass potentials are related 
through a relativistic ``third law''
\begin{eqnarray}
\d (M_1^2-M_2^2)=0,
\end{eqnarray}

\noindent
and 
\hskip 0.8cm {ii.)} they depend on the separation variable only
through the space-like projection perpendicular to the total momentum
\begin{eqnarray}
M_i=M_i(\xp),
\en
where
\bg
\xp^{\mu}=(\eta^{\mu\nu}+\Ph^{\mu}\Ph^{\nu})(x_1-x_2)_\nu.
\end{eqnarray}
Integration of  the ``third law'' condition yields 
\begin{eqnarray}
\label{eq:2250}
M_1^2-M_2^2=m_1^2-m_2^2
\end{eqnarray}
with the hyperbolic solution
\begin{eqnarray}
M_1 = m_1 \ {\rm cosh} L \ + m_2 \ {\rm sinh} L,\ 
M_2 = m_2 \ {\rm cosh} L \ + m_1 \ {\rm sinh} L,
\end{eqnarray}
given in terms of a single invariant function $L=L(\xp)$.  
\p
The $\xp$ dependence of the potential and the relativistic ``third law''
lie at the 
heart of two-body constraint dynamics. Without these conditions the constraints 
would not be compatible. While the physical importance of the $\xp$ dependence 
lies in its  exclusion of  the relative time in the c.m. frame, the 
``third law''
condition relates the mutual interactions between the
     particles to the effective potentials each particle feels in the
presence of the other in a consistent way. It is useful to show its 
implications in the simpler case of spinless particles.  
The two generalized mass shell 
constraints that are the counterparts of Eq.\ (2.22) for scalar interactions are
\bg
\c H_i=p_i^2+M_i^2 \approx 0, ~~~~~ i=1,2.
\en
The compatibility condition for these  two constraints involves the classical 
Poisson bracket
\bg
[\c H_1,\c H_2]=[p_1^2,M_2^2]+[M_1^2,p_2^2]+[M_1^2,M_2^2]\prx 0.
\en
One can see that this is satisfied provided that the ``third law'' condition 
Eq.\ (2.23) and condition (2.24) are satisfied.  (Although the ``third 
law'' solution 
Eq.\ (2.26) combined with Eq.\ (2.24) is the simplest solution, 
it is not unique $\cite{cra87}$.) 
\p
For scalar interactions  parametrized by
\bg
M_i=m_i+S_i, ~~~~i=1,2,
\en
the ``third law'' condition becomes 
\bg
m_1S_1=m_2S_2,
\en
in the nonrelativistic limit $(|S_i|<< m_i)$.  This result can also be obtained 
from Eq.\ (2.27) by keeping only terms linear in $L$.
The two constraints (2.28) can now be written as 
\bg
p_i^2+M_i^2\prx p^2+2m_iS_i+S_i^2+m_i^2-\e_i^2=0,
\en
where we have used the fact that ${\cal H}_1 - {\cal H}_2=P\cdot p \approx 0$ 
remains unchanged upon the introduction of scalar interaction in
Eq.\ (2.24).
Hence the total c.m. energy $w=\e_1+\e_2$ takes on a familiar form in the 
nonrelativistic limit
\bg
w=m_1+m_2+{p^2\ovr 2\mu} +S +O(S^2),
\en
where
\bg
S=(m_1+m_2)S_1/m_2= (m_1+m_2)S_2/m_1. 
\en

\par\nd d) The final step is to canonically quantize the classical dynamical 
system defined by 
$\c S_1$ and $\c S_2$ 
by replacing the 
Grassmann variables $\th_{\mu i},\th_{5i},\ i=1,2 $ with two mutually commuting 
sets of theta matrices, and the position and coordinate variables by operators 
satisfying the fundamental commutation relation
\begin{eqnarray}
\{x^{\mu},p^{\nu}\}\to [x^{\mu},p^{\nu}]=i\eta^{\mu\nu}.
\end{eqnarray}

The compatible pseudoclassical spin constraints $\c S_1$ and $\c S_2$ then 
become commuting quantum operators
\begin{eqnarray}
[\c S_1,\c S_2]=0.
\end{eqnarray}
The resulting CTB Dirac equations for scalar 
interactions
\bgm
\bg
\c S_1\psi = (\theta_1 \cdot p + \epsilon_1 \theta_1 \cdot \hat P + M_1
\theta_{51} - i \d L\cdot \theta_2
\theta_{52} \theta_{51})\psi=0,  
\en
\bg
\c S_2\psi = (- \theta_2 \cdot p + \epsilon_2 \theta_2 \cdot \hat P +
M_2 \theta_{52} + i \d L \cdot \theta_1
\theta_{52} \theta_{51})\psi=0,  
\en
\enm
where
\begin{eqnarray}
\d L={\d M_1\ovr M_2}={\d M_2\ovr M_1},
\end{eqnarray}
are then said to be strongly compatible. 
This strong 
compatibility has been achieved by a supersymmetry which produces the extra 
spin-dependent recoil terms involving $\d L$.
These extra terms vanish, however, when one of the particles becomes 
infinitely massive (as seen by the parametrization 
$M_i=m_i+S_i$ of  the scalar potential) 
 so that we recover the expected one-body Dirac equation in an 
external scalar potential.

Note that the Dirac constraint operators satisfy [10]
\bg
\c S_1^2-\c S_2^2=-\OOT(p_1^2+m_1^2-p_2^2-m_2^2)=-\Pp\prx 0.
\en 
Thus the relative momentum remains orthogonal to the total momentum after the 
introduction of the interaction. This also
 implies that the constituent on-shell c.m.
energies $\e_i$ are weakly equal to their off mass shell values ($-\Ph\dt p_i
\prx \e_i$). 
Notice further that since $[\Pp,M(\xp)]\sim 
P\dt\xp\equiv 0,$ 
this constraint does not violate the requirement of compatibility given 
in Eqs.\ (2.24-25).

\section{ A General Interaction for Two-Body Dirac Equations }

The previous work [16] has shown how the compatibility problem can be solved 
without having to invent new supersymmetries 
if the scalar potential is replaced by vector, pseudoscalar,
pseudovector, or tensor potentials. That work 
also relates the supersymmetric or ``external potential'' approach to the 
alternative treatment of  the two body Dirac equations of constraint dynamics 
presented by H. Sazdjian [17].  

The 
``external potential'' form Eqs.\ (2.37) of the CTB Dirac 
equations for scalar interaction can be rewritten in the hyperbolic form 
[16]
\bgm
\bg
\c S_1\psi =(\rm {cosh}\Delta~ \m S_1 +\sinh\Delta~ \m S_2)\psi=0 ,
\en
\bg
\c S_2\psi =(\cosh\Delta~ \m S_2 +\sinh\Delta~ \m S_1)\psi=0,
\en
\enm
where $\Delta$ generates the scalar potential terms in (2.37) provided that
\bg
\Delta = - \th_{51} \th_{52}  L(\xp).
\en
The operators $\m S_1$ and $\m S_2$  are auxiliary constraints of the form 
\bgm
\bg
\m S_1 \psi\equiv (\c S_{10} \cosh \Delta~ + \c S_{20} 
\sinh \Delta~) 
\psi = 0,
\en
\bg
\m S_2 \psi \equiv (\c S_{20} \cosh  \Delta~ + \c S_{10} 
\sinh \Delta~) 
\psi = 0. 
\en
\enm
To verify that the ``external potential'' forms  Eq.\ (2.37) 
result from using Eqs.\ (3.3) in Eqs.\ (3.1),
one simply commutes  the free Dirac operator $\c S_{i0}$ 
to the right onto the wave function using Eqs.\ (2.7-2.11), (2.38) and
hyperbolic identities [16].  
With this construction, the interaction enters only through an 
invariant matrix function $\Delta$ with all other spin-dependence a 
consequence of the factors contained in 
the kinetic free Dirac operators $\c S_{10}$ and $\c S_{20}$. 

Even though the form of the contraints 
Eqs.\ (3.1) and (3.3) were motivated by examining world 
scalar interactions, let us propose them for arbitrary $\D$ and 
determine their compatibility requirements. We do this for 
arbitrary  interactions by generalizing 
arguments given in Refs.[16-17]. First consider the conditions for the
compatibility of Eqs.\ (3.3a-b).
Multiplying Eq.\ (3.3a) by $\c S_{10}$ and Eq.\ (3.3b) by $\c S_{20}$ and 
subtracting we obtain 
\begin{mathletters}
\bgm
\bg
\Pp\ ({\rm cosh}\ \D)~ \psi=0.
\en
Multiplying Eq.\ (3.3b) by $\c S_{10}$ and 
Eq.\ (3.3a) by $\c S_{20}$ and 
subtracting we obtain 
\bg
\Pp\  ({\rm sinh}\ \D)~\psi=0.
\en
\enm
\end{mathletters}
We have used Eq.\ (2.2) and $\e_1-\e_2=(m_1^2-m_2^2)/w$ to simplify these 
equations.  Multiplying Eq.\ (3.4a) by ${\rm sinh}\D$, Eq.\ (3.4b) by ${\rm 
cosh}\D$,
bringing the operator $\Pp$ to the right and subtracting we find
the condition
\bg
[\Pp,\D]\psi=0.
\en
Multiplying Eq.\ (3.4a) by ${\rm cosh}\D$, Eq.\ (3.4b) by ${\rm sinh}\D$,
bringing the operator $\Pp$ to the right and subtracting we find the 
further condition
\bg
\Pp\ \psi=0.
\en
Notice that this latter condition was previously associated with the
``third law'' condition when derived from the ``external potential''
forms of the constraints (see Eq.\ (2.39)). Here the ``third law''
condition is built into the constraint by having the same generator
$\D$ for Eqs.\ (3.3a) and (3.3b).  Thus the two tentative constraints
Eqs.\ (3.3a) and (3.3b) taken together imply that for arbitrary $\D$
the orthogonality condition $\Pp\prx 0$ has to be satisfied when
acting on $\psi$. However, in order to verify that there are no
additional conditions beyond Eqs.\ (3.5) and (3.6) we must check for
mathematical consistency by examining the compatibility condition.  We
compute the commutator $[\m S_1,\m S_2]$ by rearranging its eight
terms and find that
\bg
[\m S_1,\m S_2]=[\c S_{10},{\rm cosh}\D]\m S_2-[\c S_{20},{\rm cosh}\D]\m S_1
+[\c S_{20},{\rm sinh}\D]\m S_2-[\c S_{10},{\rm sinh}\D]\m S_1
\nonumber\\
+{\rm cosh}\D(\c S_{10}^2-\c S_{20}^2){\rm sinh}\D
-{\rm sinh}\D(\c S_{10}^2-\c S_{20}^2){\rm cosh}\D
\en
does not in general vanish, unlike Eqs.\ (2.36) and (2.37).
By using Eqs.\ (2.12) and  bringing the operator $\Pp$ 
to the right, and using the conditions given in Eqs.\ (3.5) 
and (3.6) we can reduce $[\m S_1,\m S_2]\psi$ to only terms involving
$\m S_1\psi$ and $\m S_2\psi$. Since Eqs.(3.5) and (3.6) follow from combining
the constraints $\m S_i\psi=0$, 
no further conditions for mathematical consistency need be imposed on the 
constraints or the wave function.
Eq.\ (3.6) is the quantum 
counterpart of Eq.\ (2.39) but for arbitrary interactions.  Eq.\ (3.5) is also 
satisfied 
 for arbitrary  $\Delta$ provided only that the generator $\D$ satisfies  
\begin{eqnarray}
\Delta =\Delta(\xp)
\end{eqnarray}
generalizing Eq.\ (2.24). 

The weak compatibility of the ``external potential'' form of the  constraints 
Eq.\ (3.1) for general $\D$
\bg
[\c S_1,\c S_2]\psi=0
\en
can be seen by examining the four commutators
in $[\c S_1, \c S_2]$.
The commutator
\bg
[{\rm cosh}\Delta\ \m S_1,
 {\rm cosh} \Delta\ \m S_2]
= {\rm cosh} \Delta \big({\rm cosh} \Delta [\m S_1, \m S_2]
 + [\m S_1, {\rm cosh} \Delta]  \m S_2+[ {\rm cosh} 
\Delta, \m S_2] \m S_1)
\en
and is weakly zero since $\m S_i\psi=0$ and $[\m S_1,\m 
S_2]\psi=0$.
Likewise, we have
\bg
[{\rm sinh} \Delta \m S_2, {\rm sinh} \Delta \m S_1]\psi=0.
\en
The remaining two brackets are
\bg
-[{\rm cosh} \Delta\  \m S_1, {\rm sinh} \Delta\ 
\m S_1] - [{\rm sinh} \Delta\ \m 
S_2, {\rm cosh} \Delta\ \m S_2]
\nonumber\\
= [\m S_1, {\rm sinh} (\Delta)]\m S_1
-{\rm sinh}(\Delta)[{\rm cosh}(\Delta),\m S_1]\m S_1+\big(1\to 2\big).
\en
Since they contain the constraints on the right we obtain
Eq.\ (3.9) after combining with Eqs.\ (3.10) and (3.11).

One final feature should be mentioned.  Eqs.\ (3.1) and (3.3) are
also applicable  for a sum 
of the four ``polar'' interactions 
\bg
\D_1=\D_L+\D_J+\D_{\c G}+\D_{\c F},
\en
where
\begin{eqnarray}
\Delta_L= - L\th_{51}\th_{52}\ ~~~\rm {scalar,}
\end{eqnarray}
\begin{eqnarray}
\Delta_J=J\Ph\dt\th_1\Ph\dt\th_2\ ~~~{\rm {time-like}}\ 
\rm{vector,}
\end{eqnarray}
\begin{eqnarray}
\Delta_{\c G}=\c G\th_{1\perp}\dt\th_{2\perp}\ ~~~\rm{space-like}\ 
\rm{vector,}
\end{eqnarray}
\begin{eqnarray}
\Delta_{\c F} =4\c F\th_{1\perp}\dt\th_{2\perp}\th_{51}\th_{52}
\Ph\dt\th_1\Ph\dt\th_2\ ~~~\rm{tensor}\ (\rm{polar}).
\end{eqnarray}
For the sum
\bg
\D_2=\D_C+\D_H+\D_I+\D_Y
\en
of their axial counterparts
\begin{eqnarray}
\Delta_C=-C/2,\ ~~~~\rm{pseudoscalar,}
\end{eqnarray}
\begin{eqnarray}
\Delta_H =-2H\Ph\dt\th_1\Ph\dt\th_2\th_{51}\th_{52}\ 
~~~~\rm{time-like}\ 
\rm{pseudovector,}
\end{eqnarray}
\begin{eqnarray}
\Delta_I=2I\th_{1\perp}\dt\th_{2\perp}\th_{51}\th_{52}\ 
~~~~\rm{space-like}\ \rm{pseudovector,}
\end{eqnarray}
\begin{eqnarray}
\Delta_Y=-2Y\th_{1\perp}\dt\th_{2\perp}\Ph\dt\th_1\Ph\dt\th_2\ 
~~~~\rm{tensor}\ (\rm{axial}),
\end{eqnarray}
the $\sinh\D$ terms in Eq.\ (3.1) should carry negative signs instead 
[16].  
In  contrast, Eq.\ (3.3) as written remains valid as is 
for $\D_2$. For systems with both polar and 
axial interactions [16], one uses $\D_1-\D_2$ in (3.1) and $\D_1+\D_2$ in (3.3).
The terms $L,J,{\c G},{\c F},C,H,I,Y$ are arbitrary invariant functions of $\xp$. 

Eq.\ (3.3) is more convenient to use for the construction of 
Breit-like  equations 
from the constraint formalism for general interactions. Eq.\ (3.1) is more 
convenient if one aims to obtain a set of Dirac  equations in an ``external 
potential'' form similar to that exhibited in the one body Dirac equation
with the transformation properties one would expect for such
potentials.  We have already seen this for the scalar case in which
the scalar interaction ``generator'' $L$ in (3.14) in the hyperbolic
form leads to a modification of the mass term. In the case of combined
scalar, time- and space-like vector and pseudoscalar interactions, we
use the hyperbolic parametrization
\begin{mathletters}
\begin{eqnarray}
M_1 =  m_1 \ {\rm cosh}L \ + m_2~ {\rm sinh}L \,,
\end{eqnarray} 
\begin{eqnarray}
M_2 =  m_2 \ {\rm cosh}L \ + m_1~ \ {\rm sinh}L \,,
\end{eqnarray} 
\end{mathletters}
\begin{mathletters}
\begin{eqnarray}
E_1 = \e_1 \ {\rm cosh}J \ + \e_2 {\rm sinh}J  \,,
\end{eqnarray}
\begin{eqnarray}
E_2 = \e_2 \ {\rm cosh}J \ + \e_1 \ {\rm sinh}J \,,
\end{eqnarray}
\end{mathletters}
\begin{eqnarray}
G=e^{\c G},
\end{eqnarray}
where $L,J,$ and $\c G$ generate scalar, time-like vector and
space-like vector interactions respectively, while $C$ generates
pseudoscalar interactions. The resultant ``external potential'' form
for
\begin{eqnarray}
\D=\D_J+\D_L+\D_{\c G}+\D_C
\end{eqnarray}
is
\begin{mathletters}
\begin{eqnarray}
\c S_1\psi &=&\big(G \theta_1 \cdot p + E_1 \theta_1 \cdot \hat P +
M_1 \theta_{51}
\nonumber\\
& &
+ iG( \theta_2 \cdot \partial \c G \theta_{1 \perp} \dt
\theta _{2 \perp} + \theta_2 \cdot \partial J \theta_1 \cdot \hat
P \theta_2 \cdot \hat P - \theta_2 \cdot \partial L \theta_{51}
\theta_{52}+\th_2 \dt\d C/2)\big )\psi=0
\end{eqnarray}
\begin{eqnarray}
\c S_2\psi &=&  \big(- G \theta_2 \cdot p + E_2 \theta_2 \cdot \hat P + M_2
\theta_{52} 
\nonumber\\
& &
- iG (\theta_1 \cdot \partial \c  G \theta_{1 \perp} \dt
\theta_{2 \perp} + \theta_1 \cdot \partial J \theta_1 \cdot \hat
P \theta_2 \hat P - \theta_1 \cdot \partial L \theta_{51}
\theta_{52}+\th_1 \dt\partial C/2)\big )\psi = 0. 
\end{eqnarray}
\end{mathletters}
The scalar generator produces  the mass or scalar potential $M_i$
terms, the time-like vector generator produces the 
 energy or time-like potential $E_i$ terms, the space-like vector
generator
produces  the transverse or 
space-like momentum $G$ terms, while the pseudoscalar generator 
produces  only spin-dependent terms.  
Note that the vector and  scalar interactions 
also have additional spin-dependent recoil terms essential for compatibility. 
The above  features are just what one would expect for interactions 
transforming in this way and are direct consequences of the hyperbolic
parametrization of the contraints. 
Other parameterizations may not have this important property. 
	
\section{ Reduction to a Breit Equation}
We now derive a Breit equation from the CTB Dirac equations 
(3.3). Consider the combination $2(\th_1\dt\Ph\m S_1+\th_2\dt\Ph\m S_2)$.  The 
terms proportional to $w=\e_1+\e_2$ are
\bg
w[ \cosh\D+2(\th_1\dt\Ph\th_2\dt\Ph) \sinh\D]=w\ exp(\c D)
\en
where
\begin{eqnarray}
\c D=2(\th_1 \cdot \Ph~~ \th_2 \cdot \Ph) \Dlt.
\end{eqnarray}
The other terms can also be written in terms of $\c D$,  using the relations
\begin{eqnarray}
\cosh\c D=\cosh\Delta,~ \sinh\c D=2\th_1\dt\Ph\th_2\dt\Ph \sinh\Delta,
\end{eqnarray}
\begin{eqnarray}
(2\th_2\dt\Ph\th_1\dt p-2\th_1\dt\Ph\th_2\dt p)\sinh\Delta=
-(2\th_1\dt\Ph\th_1\dt p-2\th_2\dt\Ph\th_2\dt p)\sinh{\c D},
\end{eqnarray}
and
\begin{eqnarray}
(2\th_2\dt\Ph\th_{51} m_1+2\th_1\dt\Ph\th_{52} m_2)\sinh\Delta 
=-(2\th_1\dt\Ph\th_{51} m_1+2\th_2\dt\Ph\th_{52} m_2)\sinh{\c D}.
\end{eqnarray}
Then the combination $2(\th_1\dt\Ph\m S_1+\th_2\dt\Ph\m S_2)$ of the two 
hyperbolic equations in Eq.\ (3.3) takes the simple form
\begin{eqnarray}
w\ exp(\c D)\psi=(H_{10}+H_{20})exp(-\c D)\psi
\end{eqnarray}
where
\begin{eqnarray}
H_{10}=-2\th_1\dt\Ph\th_1\dt p-2\th_1\dt\Ph\th_{51} m_1
=\a_1\dt p_{\perp}+\b_1 m_1+\Ph\dt p
\end{eqnarray}
\begin{eqnarray}
H_{20}=2\th_2\dt\Ph\th_2\dt p-2\th_2\dt\Ph\th_{52} m_2
= -\a_2\dt p_{\perp}+\b_2 m_2-\Ph\dt p
\end{eqnarray}
(with the definition $p_{\perp}\equiv p+\Ph p\dt\Ph)$
are  covariant free Dirac Hamiltonians involving the covariant $\a$ and $\b$ 
matrices given previously in Eqs.\ (2.14) and (2.15).
If we take 
\begin{eqnarray}
\Psi=exp(-\c D)\psi
\end{eqnarray}
we obtain finally the manifestly covariant Breit-like equation 
\begin{eqnarray}
\label{eq:breq}
w\ exp(2\c D)\Psi=(H_{10}+H_{20})\Psi.
\end{eqnarray}

The interactions appear in the Breit equation in the exponentiated form
$exp(2\c D)$, where
$2\c D$ contains all eight interactions 
shown in $\D_1$ and $\D_2$. 
We can even add  momentum-independent 
interactions proportional to
$\s_1\dt\hat r\s_2\dt\hat r$ to $2\c D$ to give the more general interaction
\begin{eqnarray}
2\c D= && J-\b_1\b_2 L + \r_1\r_2 C + {\gm_{51}\gm_{52}H } +\s_1\dt\s_2
(-{I} + {\b_1\b_2 Y }+{ \r_1\r_2 \c F }
+\gm_{51}\gm_{52}\c G)
\nonumber\\
&&
+ {\sigma_1} \cdot \hat {{r}} { \sigma_2} \cdot \hat {{r}} ( N  
+ \beta_1 \beta_2 T 
+ \rho_1 \rho_2 S 
+  \gamma_{51} \gamma_{52} R ), 
\end{eqnarray}
where
\begin{eqnarray}
\r_i=\b_i\gm_{5i},
\end{eqnarray}
the covariant $\s$ is from Eq.\ (2.16), $\hat r = x_\perp / |x_\perp|$,  
and  $N,T,S$, and $R$ 
are arbitrary invariant functions of $\xp$.
Note that each term in $2\c D$ involves 
identical operators for particle 1 and particle 2.  As a result, they all 
commute with each other.  For example, we have
\begin{eqnarray}
[\s_1\dt\hat r\s_2\dt\hat r,~~\s_1\dt\s_2]=0\,.
\end{eqnarray}
Hence the a single exponential function $2\c D$ can also be written as
a product of separate exponentials.

Before continuing our discussion on the structure of the covariant
Breit equation, we consider the remaining linear combination of the
two constraint equations (3.3) involving the difference $
2(\th_1\dt\Ph\m S_1-\th_2\dt\Ph\m S_2)$. Using identities in the
beginning of this section we find that the difference equation becomes
\bg (\e_1-\e_2)exp(-\c D)\psi=(H_{10}-H_{20})exp(\c D)\psi \,.
\en
Transforming to $\Psi$ and using the Breit equation (4.10) gives 
\bg
(\e_1-\e_2)\Psi={(H_{10}^2-H_{20}^2)\ovr w} \Psi={(m_1^2-m_2^2)\ovr
w}\Psi +{2(H_{10}+H_{20})\Pp\ovr w}\Psi.  
\en 
Combined with Eq.\
(2.13) this gives \bg \Pp\Psi=0, \en confirming the expectation that
these momenta remain orthogonal after the interaction is introduced.
This result has also been obtained recently by Mourad and Sazdjian
[18] who emphasize that this would ensure the Poincaire' invariance of
the theory.  They further point out that the c.m. energy dependence of the
potential in the ``main equation'' (our Eq.\ (4.10)) ensures the
global charge conjugation symmetry [18] of the system, a feature that
is missing from the original Breit equation.

In summary, the constraint equations imply covariant Breit-like
equations of a certain form (4.10) whose wave function also satisfies
the constraint equation (4.16).
Alternatively, if we start off with a covariant Breit-like equation of
the form (4.10) with $\c D= \c D (\xp)$ and require simultaneously
that
$\Pp \Psi =0$, we can work backward to obtain the two compatible 
constraint equations (3.3).

We point out finally that the application 
of the constraint equations to electromagnetic
interactions does not involve the term 
$\a_1\dt\hat r\a_2\dt \hat r$ appearing in the Breit interaction 
Eq.\ (1.2). Nevertheless, it does produce the correct
spectrum, as shown in \cite{cra84,cra93}  using the ``external
potential'' form in Eqs.\ (3.27)
of the two-body Dirac equations 
with $L=C=0$, $J=-\c G$.  As has been recently noted 
[19], in the context of the Breit-like form Eq.\ (4.10)  of that 
equation, one obtains the reduction
\bg
(H_{10}+H_{20}+V_1+V_2\a_1\dt\a_2+V_3\gm_{51}\gm_{52}+V_4\s_1\dt\s_2
)\Psi = w \Psi,
\en
which contain pseudovector terms in place of 
the vector Breit term ($\a_1\dt\hat r\a_2\dt \hat r$).

\section { Structure of the Breit Equation}

The Breit equation (4.10) can be written, as usual, 
as a set of coupled equations 
for different components of the  wave function contained in $\Psi$. 
We work in the center-of-mass system for which 
$\hat P=(1, \bbox{0} )$, $\sigma=(0,\bbox{\sigma})$, and 
${\hat r}=(0,\hat {\bbox{r}} )$.  We begin by 
simplifying the general interaction (4.11) to the more compact form

\begin{eqnarray}
2{\cal D}= \sum_{\nu=0}^3 (A_\nu +  \bbox {\sigma_1} \cdot \bbox{ \sigma_2} B_\nu
 +  \bbox {\sigma_1} \cdot \hat {\bbox{r}} \, \bbox{ \sigma_2}
 \cdot \hat {\bbox{r}} \, C_\nu) q_\nu^{(1)} q_\nu^{(2)}
\end{eqnarray}
where the superscripts (1) and (2) label the interacting particles 1
and 2. 
The operators 
\begin{eqnarray}
(q_0,q_1,q_2,q_3)   = (1, \gamma_5, -i \rho, \beta) \,
\end{eqnarray}
are defined so that 
  $q_1, q_2$ and $q_3$ are analogous to 
the Pauli matrices $\sigma_1$, $\sigma_2$ and $\sigma_3$ satisfying
\begin{eqnarray}
q_i q_j = \delta_{ij} + i \epsilon_{ijk} q_k
\end{eqnarray}
where $i,j$ and $k = 1,$ 2, and 3. Eq.\ (5.1) is in the form of 
four-scalar products
\begin{eqnarray}
2{\cal D}= (A  +  \bbox {\sigma_1} \cdot \bbox{ \sigma_2} B 
 +  \bbox{ \sigma_1} \cdot \hat {\bbox{r}} \, \bbox{ \sigma_2}
 \cdot \hat {\bbox{r}} \, C ) \cdot Q 
\end{eqnarray}
involving the ``four-vector'' $Q_\nu=q_\nu^{(1)} q_\nu^{(2)}$,
and 
\bg
A=(J, H ,-C,-L),\
B=(-{I }, \c G,{- \c F },{Y}),\
C=(N,R,-S,T).
\en
 
The  wave function $\Psi$ in Eq.\ (4.10)
can be written as a spinor or column vector with two indices, one for each 
particle
\begin{eqnarray}
\Psi= \Psi^{(1)} \Psi^{(2)}.
\end{eqnarray}
It is however more 
 convenient to express the content of the wave function $\Psi$ in terms of a
new 4$\times$4 matrix wave function $\Psi'$
\begin{eqnarray}
\Psi'=\Psi^{(1)}\Psi^{(2)T} \alpha_y\,,
\end{eqnarray}
where $\Psi^{(2)}$ has been transposed and 
an operator $\a_y $ has been added on the right, as explained below.
We can represent the operation of ${\cal A}^{(1)}$ for particle 1 and ${\cal B}^{(2)}$  
for particle 2 acting on the original spinor wave function $\Psi$
in terms of operations $\cal A$ and ${\cal B}'$ 
on the new wave function $\Psi'$, 
\begin{eqnarray}
{\cal A}^{(1)} {\cal B}^{(2)} \Psi \rightarrow
{\cal A}\Psi^{(1)}[{\cal B}\Psi^{(2)}]^T\a_y
 ={\cal A}\Psi' {\cal B}',
\end{eqnarray}
where 
the matrix operator ${\c B}'$ is 
\begin{eqnarray}
{\cal B}'=\alpha_y {\cal B}^T \alpha_y\,. 
\end{eqnarray}
The arrow in Eq.\ (5.8) indicates the transformation of the operation
of ${\c A}^{(1)} {\c B}^{(2)}$ on the wave function $\Psi$ to the
operation of ${\c A}$ and ${\c B}'$ with respect to the new wave
function $\Psi'$.  The operator $\alpha_y$ in Eq.\ (5.7) insures that
operators such as $\bbox {\alpha}^{(2)}$ acting on the second particle
is represented by 
\bg \bbox{\alpha}^{(2)} \Psi \rightarrow \Psi'
\alpha_y \bbox {\alpha}^T \alpha_y = - \Psi' \bbox {\alpha} \,, 
\en
where the same negative sign appears for the different components of
the operator $\bbox{\alpha}$.  By using the wave function $\Psi'$ and
Eq.\ (5.8), an operator acting on the first particle appears on the
left side of the wave function $\Psi '$, while an operator $\c O$
acting on the second particle becomes $\alpha_y {\cal O}^T \alpha_y$
and appears on the right side of $\Psi '$.

In this matrix notation, the righthand side (RHS)  of Eq.\
({\ref{eq:breq}) becomes
\begin{eqnarray}
(H_{10}+H_{20})\Psi &\rightarrow& 
\bbox{p} \cdot \bbox{\alpha} \, \Psi' +
\bbox{p} \cdot \Psi'  \,  \bbox{\alpha}
+ m_1 \beta \Psi'
- m_2 \Psi' \beta
\equiv {\rm RHS}\,.
\end{eqnarray}
The reduction of Eq.\ ({\ref{eq:breq}) is facilitated by separating the
matrix wave function into two parts:
\begin{eqnarray}
\label{eq:psi}
\Psi=\Psi_0+\Psi_1 \rightarrow \Psi'={\Psi_0}' 
+ {\Psi_1}' \equiv \sum_{\lambda,\kappa=0}^3 q_\kappa
\sigma_\lambda \psi_{ \kappa \lambda},
\end{eqnarray}
where $\sigma_0=1$,  ${\bbox{\sigma}}=\gamma_5 {\bbox{\alpha}}$ \cite{fnsig}, 
\begin{mathletters}
\begin{eqnarray}
\Psi_0'=\sum_{\kappa=0}^3 q_\kappa \psi_{\kappa 0} , 
\end{eqnarray}
and 
\begin{eqnarray}
\Psi_1'=\sum_{\kappa=0}^3 q_\kappa  {\bbox{\sigma}} \cdot 
{\bbox \psi}_{\kappa }=
\sum_{\kappa=0}^3  \sum_{\lambda=1}^3 
q_\kappa \sigma_\lambda \psi_{\kappa \lambda}  . 
\end{eqnarray}
\end{mathletters} 
These parts give different results when operated on by 
${\bbox{\sigma}}_1 \cdot {\bbox{\sigma}}_2$,
\begin{eqnarray}
{\bbox{\sigma}}_1 \cdot {\bbox{\sigma}}_2 \, \Psi \rightarrow
-{\bbox{\sigma}} \cdot \Psi'  {\bbox{\sigma}} 
= -3 \Psi_0'+\Psi_1'.
\end{eqnarray}  
This shows that $\Psi_0 (\Psi_0')$ and $\Psi_1(\Psi_1')$ 
are respectively the
spin-singlet and spin-triplet parts of the wave function.
Furthermore, all the operators in Eq.\ (4.11) commute.  With 
$({\bbox{\sigma}}_1 \cdot \hat {\bbox{r}} \, 
{\bbox{\sigma}}_2 \cdot \hat {\bbox{r}})^2=1$, 
the tensor interaction can be written as 
\begin{eqnarray} 
\label{eq:ten}
e^{  {\bbox{\sigma}}_1 \cdot \hat {\bbox{r}} \, 
{\bbox{\sigma}}_2 \cdot \hat {\bbox{r}} \, C \cdot 
Q} 
=
\cosh ( C \cdot Q) 
+ 
{\bbox{\sigma}}_1 \cdot \hat {\bbox{r}} \, {\bbox{\sigma}}_2 \cdot \hat 
{\bbox{r}} \,
\sinh(  C \cdot Q ) \,.
\end{eqnarray}
This means that interaction (4.11) alone will not mix spin-singlet and
spin-triplet states (as seen below in Eq.(5.34), the kinetic 
energy term does mix singlet and triplet states).

The reduction to the matrix form is easier for the spin-singlet wave
function\break 
$\Psi_0={1\over 4} (1 - {\bbox{\sigma}}_1 \cdot {\bbox{\sigma}}_2)
\Psi$ because of its simpler spin structure:
\begin{eqnarray}
e^{( A +  \bbox{\sigma_1} \cdot \bbox{\sigma_2} B 
+ {\bbox{\sigma}}_1 \cdot \hat {\bbox{r}}\,  {\bbox{\sigma}}_2 \cdot \hat
 {\bbox{r}} \,
C) \cdot Q } \Psi_0  = e^{T\cdot Q} \Psi_0, 
\end{eqnarray}
where $T=A-3B-C$.  A Taylor expansion of the expontial operator on the
above equation shows that it is necessary to evaluate the basic
operation of the type
\begin{eqnarray} 
T \cdot Q \Psi_0 = \sum_{\nu=0}^3 T_\nu q_\nu^{(1)} q_\nu^{(2)}
\Psi_0.
\end{eqnarray}
In terms of the new wave function $\Psi_0'$ of
Eq.\ (5.13a), the above equation can be represented in the matrix form:
\begin{eqnarray} 
\sum_{\kappa=0}^3 \sum_{\nu=0}^3 
T_\nu (  q_\nu q_\kappa q_\nu' )\psi_{\kappa 0}
= \sum_{\kappa=0}^3 T \cdot S_\kappa q_\kappa  \psi_{\kappa 0}\,,
\end{eqnarray}
where $q_\nu '=\a_yq_{\nu}^T\a_y=\epsilon_\nu q_\nu$, with
$( \epsilon_0, \epsilon_1, \epsilon_2, \epsilon_3)=(1,1,1,-1)$ and
we have introduce the quantity $(S_\kappa)_\nu$ defined by 
\begin{mathletters}
\begin{eqnarray}
\label{eq:qqq}
q_\nu q_\kappa q_\nu' = (S_\kappa)_\nu q_\kappa,
\end{eqnarray}
and
\begin{eqnarray}
(S_\kappa)_\nu=\epsilon_\nu[1+2(1-\delta_\kappa 0)( 1-\delta_{\nu 0})
(\delta_{\kappa \nu} -1)]\,.
\end{eqnarray}
\end{mathletters}
The $4\times4$ matrix $S$ is called a signature matrix because its
matrix elements can only be $+1$ or $-1$.  The row vectors of $S$ are
\begin{eqnarray}
\label{eq:sk1}
S_{ 0  } &=& ( 1, ~1, ~1, -1),\nonumber\\
S_{ 1  } &=& ( 1, ~1, -1, ~1), \nonumber\\
S_{ 2  } &=& ( 1, -1, ~1, ~1), \nonumber\\
S_{ 3  } &=& ( 1, -1, -1, -1) \,.
\end{eqnarray}
Eq.\ (5.18) can be applied repeatedly to give the desired result
\begin{eqnarray}
w e^{T \cdot Q} \Psi_0 &=& w \sum_{n=0}^\infty {1 \over n!}(T\cdot
Q)^n \Psi_0 \nonumber\\
&\rightarrow& w \sum_{\kappa=0}^3 e^{T\cdot S_\kappa} q_\kappa
\psi_{\kappa 0} \equiv {\rm LHS}_0. 
\end{eqnarray}
It can also be used to prove the general result
\begin{eqnarray}
f({T \cdot Q}) \Psi_0 
\rightarrow  \sum_{\kappa=0}^3 f({T\cdot S_\kappa}) q_\kappa
\psi_{\kappa 0}. 
\end{eqnarray}
 
The treatment of the spin-triplet expression 
\begin{eqnarray} 
e^{2{\cal D}}
\Psi_1
=
e^{({A+B})\cdot Q} [\cosh ( C \cdot Q) 
+ 
{\bbox{\sigma}}_1 \cdot \hat {\bbox{r}} \, {\bbox{\sigma}}_2 \cdot \hat {\bbox{r}} \,
\sinh(  C \cdot Q )]\Psi_1 
\end{eqnarray}
is simplified by noting that the $q_\kappa$ and $\bbox{\sigma}$
matrices are independent of each other.  Hence the $Q$ dependences
can be eliminated in favor of the signature vector $S_\kappa$ with the
help of Eq.\ ({\ref{eq:qqq}), which also applies to the $q$ structure of
$\Psi_1$.  
This leaves the spin-dependent part which has the form 
\begin{eqnarray} 
{\bbox\sigma}_1 \cdot {\hat{\bbox{r}}} \, 
{\bbox\sigma}_2 \cdot {\hat{\bbox{r}}} \Psi_1 
&\rightarrow& - \bbox\sigma \cdot 
{\hat{\bbox{r}}} {\Psi_1}' \bbox\sigma \cdot {\hat{\bbox{r}}} 
\nonumber\\
&=& \sum_{\kappa=0}^3 q_\kappa (\bbox{\sigma} \cdot {\bbox{\psi}}_\kappa
-2 \bbox{\sigma} \cdot {\hat{\bbox{r}}} \, 
{\hat{\bbox{r}}}\cdot \bbox{\psi_\kappa} )\,.
\end{eqnarray}
Hence, using Eq.\ ({\ref{eq:ten}),
the spin-triplet part of the left hand side of Eq.\ (4.10) is
\begin{eqnarray}
{\rm LHS}_1=
 w \sum_{\kappa=0}^3 e^{(A+B) \cdot S_\kappa} 
q_\kappa [ e^{C\cdot S_\kappa}
\bbox{\sigma} \cdot \bbox{\psi}_\kappa
- 2 \sinh ( C \cdot S_\kappa) 
\bbox{\sigma} \cdot {\hat{\bbox{r}}} \, 
{\hat{\bbox{r}}}\cdot \bbox{\psi}_{\kappa}]\,.
\end{eqnarray}
The Breit equation (4.10) for the matrix wave function is thus 
\begin{eqnarray}
{\rm LHS}_0+{\rm LHS}_1= {\rm RHS}
\end{eqnarray}
where the expressions are from 
Eqs.\ (5.11), (5.21), and (5.25).  

Eq.\ (5.26) can be written explicitly as
$$
\sum_{\kappa \lambda=0}^3 q_\kappa \sigma_\lambda \biggl \{ 
W_{\kappa \lambda} \psi_{\kappa\lambda}
-(1-\delta_{\lambda 0}) {\bar V}_\kappa {\hat r}_\lambda 
\sum_{\mu=1}^3 {\hat r}_{ \mu} 
     \psi_{\kappa \mu} \biggr \}
~~~~~~~~~~~~~~~~~~~~~~~~~~~~~~~~~~~~~~~~~~~~~~~$$
\begin{eqnarray}
~~~~~~~~~~~
= \sum_{\kappa \lambda=0}^3 ~\biggl [ 
\sum_{i=1}^3 \biggl \{ \alpha_i q_\kappa \sigma_\lambda + q_\kappa
\sigma_\lambda \alpha_i \biggr \} p_i 
+ m_1 q_3 q_\kappa \sigma_\lambda
- m_2  q_\kappa \sigma_\lambda q_3 \biggr ] \psi_{\kappa \lambda} \,,
\end{eqnarray}
where
\begin{mathletters}
\begin{eqnarray}
W_{\kappa \lambda}= 
w\ e^{(A-3B-C)\cdot S_\kappa} \delta_{\lambda 0}
+w\ ( 1 - \delta_{\lambda 0}) e^{(A+B+C)\cdot S_\kappa}\,,
\end{eqnarray}
\begin{eqnarray}
{\bar V}_{\kappa }= 
2w e^{(A+B)\cdot S_\kappa}
\sinh (C\cdot S_\kappa) \,. 
\end{eqnarray}
\end{mathletters}
Multiplying the equation from the right by $\sigma_\lambda q_\kappa$ and
taking traces we finally get 
a set of 16 coupled equations for the wave function
components,
$$\biggl \{ W_{\kappa \lambda} \psi_{\kappa \lambda} \delta_{\lambda 0}
-(1-\delta_{\lambda 0}) {\bar V}_\kappa  {\hat r}_\lambda 
\sum_{\mu=1}^3  {\hat r}_\mu 
     \psi_{\kappa \mu}  \biggr \}
~~~~~~~~~~~~~~~~~~~~~~~~~~~~~~~~~~~~~~~~~~~~~~$$
\begin{eqnarray}
~~~~~
= \sum_{\kappa' \lambda'=0}^3 
\sum_{i=1}^3 (1+f_{\kappa' 1} f_{\lambda' i})
g_{\kappa 1 \kappa'} g_{\lambda 1 \lambda'} p_i \psi_{\kappa'
\lambda'}
+\sum_{\kappa' =0}^3 (m_1 - f_{\kappa' 3} m_2) g_{\kappa 3 \kappa'}
\psi_{\kappa ' \lambda}\,,
\end{eqnarray}
where
\begin{eqnarray}
f_{\kappa i} = \delta_{\kappa 0} +(1-\delta_{\kappa 0}) ( 2
\delta_{\kappa i} -1)\,,
\end{eqnarray}
and 
\begin{eqnarray}
g_{\kappa i \kappa'} = \delta_{\kappa 0} \delta_{ i \kappa'}
+\delta_{\kappa' 0} \delta_{ i \kappa} 
+ (1-\delta_{\kappa 0}) (1-\delta_{\kappa' 0}) i \epsilon_{\kappa i
\kappa'}
\,. 
\end{eqnarray}
The structure of these equations becomes more transparent by writing 
them out explicity in terms of the singlet and triplet wave functions 
\begin{eqnarray}
(\psi_{00}, \psi_{10}, \psi_{20},
\psi_{30}) 
=(\psi, \phi, i \chi, \eta),
\end{eqnarray}
\begin{eqnarray}
(\bbox{\psi}_{0 },\bbox{\psi}_{1 },
\bbox{\psi}_{2  },
\bbox{\psi}_{3  })
=
(\bbox{\psi}, \bbox{\phi}, i \bbox{\chi}, 
\bbox{\eta})
\,:
\end{eqnarray}
\begin{mathletters}
\begin{eqnarray} 
[w-U_{\psi}]\psi=2\bbox{p} \dt\bbox{\phi}+ (m_1-m_2)\eta,
\end{eqnarray}
\begin{eqnarray} 
[w-U_{\phi}]\phi=2\bbox{p} \dt\bbox{\psi}+ (m_1+m_2)\chi,
\end{eqnarray}
\begin{eqnarray} 
[w-U_{\chi}]\chi= (m_1+m_2)\phi,
\end{eqnarray}
\begin{eqnarray} 
[w-U_{\eta}]\eta= (m_1-m_2)\psi,
\end{eqnarray}
\begin{eqnarray} 
[w-V_{\bbox{\psi}}-\bar V_{\bbox{\psi}}\hat r\hat r\dt]\bbox{\psi}
=2\bbox{p}\phi+(m_1-m_2)\bbox{\eta}, 
\end{eqnarray}
\begin{eqnarray} 
[w-V_{\bbox{\phi}}-\bar V_{\bbox{\phi}}\hat r\hat r\dt]\bbox{\phi}=2\bbox{p}\psi+(m_1+m_2)\bbox{\chi},
\end{eqnarray}
\begin{eqnarray} 
[w-V_{\bbox{\chi}}-{\bar V}_{\bbox{\chi}}\hat r\hat r\dt]\bbox{\chi}
=-2i\bbox{p}\times\bbox{\eta}+(m_1+m_2)\bbox{\phi},  
\end{eqnarray}
\begin{eqnarray} 
[w-V_{\bbox{\eta}}-\bar V_{\bbox{\eta}}\hat r\hat r\dt]\bbox{\eta}
=-2i\bbox{p}\times\bbox{\chi}+(m_1-m_2)\bbox{\psi},
\end{eqnarray}
\end{mathletters}
where
\begin{mathletters}
\begin{eqnarray} 
w-U_{\k}=we^{(A-3B-C)\dt S_\kappa},
\end{eqnarray}
\begin{eqnarray} 
w-V_{\k}=we^{(A+B+C)\dt S_\kappa},
\end{eqnarray}
\end{mathletters}
and $\bar V_{\k}$ is defined in Eq.\ (5.28).

In order to see explicitly the distinction between the traditional Breit 
approach and that of constraint dynamics, we perform
an angular momentum decomposition.  
For the spin-zero wave functions we take
\bgm
\bg
\psi=\psi_{j} Y_{jm},\ \phi=\phi_{j} Y_{jm},\ \chi=\chi_{j} Y_{jm},\
\eta=\eta_{j} Y_{jm},
\en
\enm
where $ Y_{jm}$ is an ordinary spherical harmonic ($j=l$ here).
For the spin-one wave functions, we take a form that depends on the 
spatial parity:
\bgm
\bg
\bbox{\phi}=\phi_{jy}\bbox{ Y_{jm}}+\phi_{jz}\bbox{Z_{jm}},
\ {\rm or} \ =\phi_{jx}\bbox{X_{jm}};
\en
\bg
\bbox{\psi}=\psi_{jy}\bbox{Y_{jm}}+\psi_{jz}\bbox{Z_{jm}},
\ {\rm or} \ =\psi_{jx}\bbox{X_{jm}};
\en
\bg
\bbox{\eta}=\eta_{jy}\bbox{Y_{jm}}+\eta_{jz}\bbox{Z_{jm}},
\ {\rm or} \ =\eta_{jx}\bbox{X_{jm}};
\en
\bg
\bbox{\chi}=\chi_{jy}\bbox{Y_{jm}}+\chi_{jz}\bbox{Z_{jm}},
\ {\rm or} \ =\chi_{jx}\bbox{X_{jm}};
\en
\enm
where the three orthonormal vector spherical harmonics are
\bg
\bbox{X_{jm}}={\bbox{L}\ovr \sqrt{j(j+1)}}Y_{jm},\ 
\bbox{Y_{jm}}={\bbox{r}\ovr r}Y_{jm},\ 
\bbox{Z_{jm}}=i{r\bbox{p}\ovr \sqrt{j(j+1)}}Y_{jm} \,.
\en
The first and last vanish for $j=0$.
\p
To obtain the radial wave equations, we use
the following identities
\bgm
\bg
{\bbox{r} \over r} \dt\bbox{X_{jm}}=0,\  \bbox{p}\dt\bbox{X_{jm}}=0,
\en
\bg
{\bbox{r}\ovr r}\times \bbox{X_{jm}}=i\bbox{Z_{jm}},\ 
\bbox{p}\times \bbox{X_{jm}}={\sqrt{j(j+1)}\ovr r}\bbox{Y_{jm}}
+{\bbox{Z_{jm}}\ovr r},
\en
\bg
{\bbox{r} \over r} \dt\bbox{Y_{jm}}= Y_{jm},\ 
\bbox{p}\dt\bbox{Y_{jm}}=-{2i\ovr r} Y_{jm},
\en
\bg
{\bbox{r} \over r} \times\bbox{Y_{jm}}=0,\ 
\bbox{p}\times\bbox{Y_{jm}}=-{\sqrt{j(j+1)}\ovr r} \bbox{X_{jm}},
\en
\bg
{\bbox{r} \over r} \dt\bbox{Z_{jm}}=0,\ 
\bbox{p}\dt\bbox{Z_{jm}}=i{\sqrt{j(j+1)}\ovr r} Y_{jm},
\en
\bg
{\bbox{r} \over r} \times\bbox{Z_{jm}}=\bbox{X_{jm}},\ 
\bbox{p}\times\bbox{Z_{jm}}={\bbox{X_{jm}}\ovr r}. 
\en
\enm
The equations so obtained can be divided into two sets of different 
total parity. One set has the natural parity solution:
\bg
\c P\Psi_{jm}=(-)^{j+1}\Psi_{jm},
\en
and involves the 8 wave functions: 
\bgm
\bg
\phi=\phi_{j} Y_{jm},\ \chi=\chi_{j} Y_{jm},
\en
\bg
\bbox{\phi}=\phi_{jx}\bbox{X_{jm}},\ 
\bbox{\psi}=i\psi_{jy}\bbox{Y_{jm}}+i\psi_{jz}\bbox{Z_{jm}},
\en
\bg
\bbox{\chi}=\chi_{jx}\bbox{X_{jm}},\ 
\bbox{\eta}=i\eta_{jy}\bbox{Y_{jm}}+i\eta_{jz}\bbox{Z_{jm}}.
\en
\enm
(Four of the wave functions
$\phi_{jx},\psi_{jz},\chi_{jx}$ and $\eta_{jz}$ do not appear for $j=0$, 
because their angular parts vanish.) 
The eight simultaneous equations satisfied by them are
\bgm
\bg
(w-U_{\phi})\phi_j=2\psi\pri_{jy}+{4\ovr r}\psi_{jy}-2{\sqrt{j(j+1)}\ovr 
r}\psi_{jz}+(m_1+m_2)\chi_j,
\en
\bg
(w-V_{\bbox{\psi}}-\bar V_{\bbox{\psi}})\psi_{jy}=-2\phi\pri_j
+(m_1-m_2)\eta_{jy},
\en
\bg
(w-V_{\bbox{\eta}})\eta_{jz}=-2\chi\pri_{jx}-{2\ovr r}\chi_{jx}
+(m_1-m_2)\psi_{jz},
\en
\bg
(w-V_{\bbox{\chi}})\chi_{jx}=2\eta\pri_{jz}+{2\ovr r}\eta_{jz}
-2{\sqrt{j(j+1)}\ovr r}\eta_{jy}+(m_1+m_2)\phi_{jx}.
\en
\enm
\bgm
\bg
(w-U_\chi)\chi_j=(m_1+m_2)\phi_j,
\en
\bg
(w-V_{\bbox{\psi}})\psi_{jz}
=-2{\sqrt{j(j+1)}\ovr r}\phi_j+(m_1-m_2)\eta_{jz},
\en
\bg
(w-V_{\bbox{\phi}})\phi_{jx}=(m_1+m_2)\chi_{jx},
\en
\bg
(w-V_{\bbox{\eta}}-\bar V_{\bbox{\eta}})\eta_{jy}
=-2{\sqrt{j(j+1)}\ovr r}\chi_{jx}+(m_1-m_2)\psi_{jy} \,.
\en
\enm
The second set has the ``unnatural'' parity 
\bg
\c P\Psi_{jm}=(-)^{j}\Psi_{jm},
\en
and involves the 8 wave functions: 
\bgm
\bg
\psi=\psi_{j} Y_{jm},\ \eta=\eta_{j} Y_{jm},
\en
\bg
\bbox{\psi}=\psi_{jx}\bbox{X_{jm}},\ 
\bbox{\phi}=i\phi_{jy}\bbox{Y_{jm}}+i\phi_{jz}\bbox{Z_{jm}},
\en
\bg
\bbox{\eta}=\eta_{jx}\bbox{X_{jm}},\ 
\bbox{\chi}=i\chi_{jy}\bbox{Y_{jm}}+i\chi_{jz}\bbox{Z_{jm}}.
\en
\enm
The eight simultaneous equations satisfied by them are
\bgm
\bg
(w-U_{\psi})\psi_j=2\phi\pri_{jy}+{4\ovr r}\phi_{jy}-2{\sqrt{j(j+1)}\ovr 
r}\phi_{jz}+(m_1-m_2)\eta_j,
\en
\bg
(w-V_{\bbox{\chi}})\chi_{jz}=-2\eta\pri_{jx}-{2\ovr r}\eta_{jx}
+(m_1+m_2)\phi_{jz},
\en
\bg
(w-V_{\bbox{\eta}})\eta_{jx}=2\chi\pri_{jz}+{2\ovr r}\chi_{jz}
-2{\sqrt{j(j+1)}\ovr r}\chi_{jy}+(m_1-m_2)\psi_{jx}.
\en
\bg
(w-V_{\bbox{\phi}}-\bar V_{\bbox{\phi}})\phi_{jy}
=-2\psi\pri_j+ (m_1+m_2)\chi_{jy},
\en
\enm
\bgm
\bg
(w-U_\eta)\eta_j=(m_1-m_2)\psi_j,
\en
\bg
(w-V_{\bbox{\psi}})\psi_{jx}=(m_1-m_2)\eta_{jx},
\en
\bg
(w-V_{\bbox{\phi}})\phi_{jz}=-2{\sqrt{j(j+1)}\ovr r}\psi_j
+ (m_1+m_2)\chi_{jz},
\en
\bg
(w-V_{\bbox{\chi}}-\bar V_{\bbox{\chi}})\chi_{jy}
=-2{\sqrt{j(j+1)}\ovr r}\eta_{jx}+ (m_1+m_2)\phi_{jy},
\en
\enm
where the terms $w-U$, $w-V$ are given in Eqs.\ (5.35a-b) and $w-V-\bar V$ 
by
\bg
w-V_{\k}-\bar V_{\k}=w e^{(A+B-C)\dt S_{\k}}.
\en

Each set of 8 equations consists of 4 differential equations [Eqs.\
(5.42) or (5.46)] and 4 algebraic equations [Eqs.\ (5.43)
or (5.47)]. The algebraic equations can be used to express
4 of the 8 wave functions in terms of the remaining 4.  For
example, two of the eliminated wave functions are

\bg 
\chi_j={m_1+m_2\ovr w-U_{\chi}}\phi_j \,, 
\en
\bg  
\eta_j={m_1-m_2\ovr w-U_{\eta}}\psi_j\ \,. 
\en 
After the elimination, pole singularities
could appear in the differential equations at particle separations $r$
where equations such as Eq.\ (5.49) have poles, e.g., where 
\bg
w-U_{\chi}=0;\ w-U_{\eta}=0,\ m_1\not = m_2 \en 
provided that the total
relativistic center-of-mass energy $w$ is nonzero.  These are the
well-known singularities that plague the traditional Breit equations.

              However, from the perspective of the exponentiated
interactions of constraint dynamics such as that shown in Eq.\ (5.35),
these structural poles can appear only if the potential generators in
the exponent go to $-\inf$. In the absence of such singular behavior,
the constraint two-body equations, or their Breit analogs, are free of
the pathologies described in [6,7]. These pathologies arise because
the wave functions appearing on the right-hand side of Eq.\ (5.49) must
have zeros at the pole positions. These additional boundary conditions
give rise to spurious resonances in the continuum for sufficiently
strong interactions, and to spurious bound states for any nonzero
interaction strength at total energies (including rest masses) which
go to zero.

              We should add, for the sake of completeness, that
another class of spurious solutions appears when the values of the
exponential generators go to $+\inf$. Then the algebraic equations
impose the new boundary conditions that the wave functions appearing
on the left-hand sides of these equations must vanish at these
singular points.

It is important to point out that if the exponentiated potentials given 
in Eqs.\ (5.35) and (5.48)  are 
approximated by finite Taylor approximants, it might be possible to satisfy the 
singularity conditions such as (5.50) whenever the approximant has a zero  
where the true value of the exponential function is nonzero.  In other words, 
it is the exponential structure of the effective interaction that protects the 
CTB Dirac equations from the undesirable pathology.
We trace that exponential structure to the hyperbolic forms (3.3) of the 
constraint equations which in turn are motivated by the requirements
of compatibility. Recall that in those forms the ``generators'' $\D$ 
produce potential terms in the ``external potential'' form of the  constraint
equations which have the expected transformation properties, including the 
essential recoil corrections.

Our result can also be interpreted in another way.  The exponential functions 
appearing on the left hand side of Eq.\ (4.10) can be expanded in terms of 
hyperbolic functions. Indeed, one could express the ``effective potentials'' 
contained in $exp(2\c D)$ in the form given by the right-hand side of 
Eq.\ (4.11). If we now parametrize these 
effective potentials directly using functions with
no singularity at finite $r$, we will find that under favorable circumstances 
the singularity condition (5.50) can still be satisfied.  This is in fact the 
result  of [6,7].  Thus the regularity of the effective potentials 
appearing in $exp(2\c D)$ does not guarantee the regularity of the resulting 
Breit equation.  It is the regularity of the basic potentials appearing in 
the exponent $2\c D$ that guarantees the regularity of the resulting Breit 
equation.

\section{ Comparison of the Constraint and Breit Equations for QED}

In this section we discuss the implications of our new Breit-like
equation in quantum electrodynamics. Consider first the original Breit
equation whose matrix form is \bg (w+{\a\ovr r})\Psi'+a{\a\ovr
r}\bbox\a\dt\Psi'\bbox\a +b{\a\ovr r}\bbox\a\dt\rh\Psi'\bbox\a\dt\rh =
\bbox p\dt\bbox\a \Psi'+ \bbox
{p}\dt\Psi'\bbox\a+m_1\beta\Psi'-m_2\Psi'\beta \,, \en 
where $a$ and $b$ are potential parameters. By identifying it
with the center-of-mass form of the Breit-like form Eq.\ (4.10) of our
covariant constraint equation, we can solve for the generators given
by Eq.\ (5.5).  Using Eqs.\ (5.26) or equivalently Eqs.\ (5.27-28) we
obtain the following twelve equations for the potentials.  
\bgm \bg
exp[(A-3B-C)\dt S_0]=1+(1+3a+b)\zeta, 
\en 
\bg exp[(A-3B-C)\dt
S_1]=1+(1+3a+b)\zeta, 
\en 
\bg exp[(A-3B-C)\dt S_2]=1+(1-3a-b)\zeta,
\en 
\bg exp[(A-3B-C)\dt S_3]=1+(1-3a-b)\zeta, 
\en 
\bg exp[(A+B+C)\dt
S_0]=1+(1-a-b)\zeta, 
\en 
\bg exp[(A+B+C)\dt S_1]=1+(1-a-b)\zeta, 
\en
\bg exp[(A+B+C)\dt S_2]=1+(1+a+b)\zeta, 
\en 
\bg exp[(A+B+C)\dt
S_3]=1+(1+a+b)\zeta, 
\en 
\bg exp[(A+B-C)\dt S_0]=1+(1-a+b)\zeta, 
\en
\bg exp[(A+B-C)\dt S_1]=1+(1-a+b)\zeta, 
\en 
\bg exp[(A+B-C)\dt
S_2]=1+(1+a-b)\zeta, 
\en 
\bg exp[(A+B-C)\dt S_3]=1+(1+a-b)\zeta, 
\en
\enm 
in which $\zeta=\a/(wr)$.  The first four equations come from the
singlet wave function ($\l=0$ terms in Eq.(5,28a)),
while the last eight arise from the triplet,
with the last four coming from a combination of Eq.\ (5.28b) and the
triplet (or $\l\not =0$) part of Eq.\ (5.28a).  
These twelve algebraic equations can be
solved for the twelve unknown generators shown in Eq.\ (5.5). One can
readily show that six of these generators vanish (corresponding to
scalar, pseudoscalar, and tensor interactions): \bg S=T=\c F=Y=C=L=0.
\en The remaining generators, $J, H, I, G, N$ and $R$, are nonzero,
but we shall not need them in our discussion except in the
combinations appearing on the left-hand sides of the algebraic
equations, Eqs.\  (5.43) and (5.47). These are just the six equations
shown in Eqs.\  (6.2c-f) and (6.2k,l), now expressible as 
\bg
w-U_{\chi}=w-U_{\eta}=w~ exp[J+3I-N-H+3\c G+R]=1+(1-3a-b)\zeta \,,
\en 
\bg
w-V_{\bbox\psi}=w-V_{\bbox\phi}=w~ exp[J-I+N+H+\c G+R]=1+(1-a-b)\zeta\,,
\en 
\bg w-V_{\bbox\chi}-\bar V_{\bbox\chi}=w-V_{\bbox\eta}-\bar
V_{\bbox\eta}=w~ exp[J-I+N-H+\c G-R]=1+(1+a-b)\zeta \,.
\en 
Singularities
at finite particle separation $r$ arise, for nonzero $w$, when the
right-hand sides of these equations vanish (assuming that the
numerators in the algebraic equations in which these occur do
not). Recalling the context in which these equations appear, we see
that this occurs when \bg 3a+b>1 \en for all $\c P=(-)^{j+1}$ states,
and all $\c P=(-)^{j}$ states when $m_1\not =m_2$; \bg a+b>1 \en for
all $\c P=(-)^{j}$ states, all $\c P=(-)^{j+1}$ states when $m_1\not
=m_2$; and for $j=0$ states when $m_1=m_2$ and \bg b-a>1 \en for all
$\c P=(-)^{j+1}$ states when $m_1\not =m_2$, and for $j=0$ states when
$m_1=m_2$ and all $\c P=(-)^{j}$ states.  This confirms the result
first derived in Ref.\ [7].

It is worth noting that the original Breit equation 
corresponds to $a=b=1/2$ while the Barut equation corresponds to 
$a=1,b=0$ [21].  Both equations are therefore singular 
according to the first of the above three conditions.  

From the perspective of the exponential generators of constraint
dynamics, the vanishing of the right-hand sides of Eqs.\ (6.4-6.6) is
possible only when the generators take on the value $-\inf$ at finite
$r$. Hence they can be avoided by simply using generators not having
such $r$ singularities.  Indeed, in their constraint approach to QED,
Crater and Van Alstine [11,12] found that 
\bg e^{\c
G}=e^{-J}=(1+2\a/(wr))^{-1/2} 
\en 
with all of the remaining ten
generators set equal to zero.  As a result, no midrange zero appears 
since
\bg w-U_{\chi}=w~\exp(J+3\c G)=w~\exp(2\c G)={1\ovr
1+2\a/wr}
\en 
is finite for positive $r$.  Furthermore, the Lorentz
nature of the two nonzero generators matches that of the vector
character of the QED interaction at this level.  In contrast, the
Breit generators involve pseudovector as well as vector parts and
additional vector and pseudovector ``tensor terms'' corresponding to
the Coulomb gauge ( as opposed to the Feynman gauge used in the
constraint equations).

The conclusions obtained in this section for QED can be extended 
to other interactions (scalar, pseudoscalar, psuedovector and tensor)
important in semiphenomenological applications in nuclear and particle 
physics.  Thus, one may with safety solve the new Breit equations 
numerically with no concern about structural singularities which would 
otherwise render such solutions meaningless.

\section{ Conclusions and Discussions}

    Breit equations and constraint Dirac equations of relativistic
two-body quantum mechanics have markedly different properties: Breit
equations could have structural singularities at finite particle
separations even when the interactions themselves are nonsingular
there, while constraint Dirac equations seem to be free of
them. CTB Dirac equations are manifestly covariant, while Breit
equations are not, being valid only in the center-of-mass frame. We
are able to understand, and to reconcile, their differences in this
paper.

    The constraint Dirac equations were originally derived for scalar
interactions with the help of supersymmetries in addition to Dirac's
constraint dynamics. Generalizing this concept to arbitrary interactions
leads to a ``hyperbolic'' form of the equations. The two single-particle
Dirac equations can then be recast into two other equations: (1) a covariant
Breit-like equation with exponentiated interactions, and (2) a covariant
equation describing an additional orthogonality condition on the 
relative momentum.  We use the equivalence between these two types of 
equations to show
that the constraint Dirac equations are completely free of the unphysical
structural singularities when the exponential structure of the
interactions are not tampered with.

    The advantage of the constraint form of the Breit equation is that the
structural singularities of the traditional Breit equation are now moved
entirely to the exponential generators of the interaction. As a
consequence, they can be eliminated right from the beginning by the simple
requirement that these generators themselves be free of singularities at
finite separations. The resulting Breit equations are then guaranteed to
be free of the undesirable structural singularities that plague
traditional Breit equations. These improved Breit equations, which are
dynamically equivalent to the constraint Dirac equations, can now be used
in nonperturbative descriptions of highly relativistic and
strong-field problems such as those
appearing in two-body problems in
quantum electro- and chromo-dynamics, an in nucleon-nucleon scattering.

Of course, the constraint Dirac equations can also be used, now
that we know how to keep them singularity-free. However, in actual
applications, they have to be reduced down to a set of coupled
differential equations before actual solutions can be attempted. These
differential equations are transforms (see Eq.(4.9)) of those obtained from 
the equivalent Breit equations. So in reality the two different formulations
have now become completely identical to each other. 

{\bf Acknowledgement:}
This research was supported by the Division of Nuclear
Physics, U.S. Department of Energy under Contract
No. DE-AC05-84OR21400 
managed by 
Lockheed Martin Energy Systems.  One of the authors (HWC) wishes to
acknowledge very useful discussions with P. Van Alstine, M. Moshinsky and 
A. Del Sol Mesa on closely related topics

\vfill\eject

\vfill\eject

\begin{references}

\bibitem{brt29}
G. Breit, Phys. Rev. {\bf 34}, 553 (1929)

\bibitem{kfy38}


N. Kemmer, Helv. Phys. Act. 10, 48 (1937),
E. Fermi, and C. N. Yang, Phys. Rev. 76, 1739 (1949),
H. M. Mosley and N. Rosen, Phy. Rev. 80, 177 (1950).


\bibitem{bet57}
H.A. Bethe and E. E. Salpeter,  { \it Quantum Mechanics of One and} 
\nd
{\it Two Electron Atoms} (Springer, Berlin, 1957).

\bibitem{kro81} W. Krolikowski, Acta Physica Polonica, {\bf B12}, 891
(1981).  Nonperturbative treatments of truncated version of the Breit
equation (with just the Coulomb term) have yielded the same
results as a perturbative treatment of the same truncations but none
of these have included the troublesome transverse photon parts. See J.
Malenfant, Phys. Rev. A {\bf 43}, 1233 (1991), and T.C. Scott,
J. Shertzer, and R.A. Moore, ibid \uline{45},4393 (1992)

\bibitem{chi82}
R. W. Childers, Phys. Rev. D26, 2902 (1982).

\bibitem{won93a} 
C. W. Wong and C. Y. Wong
Phys. Lett. B301, 1 (1993).

\bibitem{won93b} 
C. W. Wong and C. Y. Wong
Nucl. Phys. A562,  598 (1993).

\bibitem{ko68}
Y. Koide, Prog. Theo. Phys. 39, 817 (1968); Nuovo Cim. 70, 411 (1982).

\bibitem{di64}
P.A.M. Dirac, \uline{Lectures on Quantum Mechanics} (Yeshiva University, Hew 
York, 1964).

\bibitem{cra82}
P. Van Alstine and H.W. Crater, J. Math. Phys. \uline{23}, 1997 (1982)
H. W. Crater and P. Van Alstine, Ann. Phys. (N.Y.) \uline{148}, 57 (1983).

\bibitem{cra84}
H. W. Crater and P. Van Alstine, 
Phys. Rev. Lett. \uline {53}, 1577 (1984), 

P. Van Alstine and H. W. Crater, Phys. Rev. \uline{D34}, 
1932 (1986).

H. W. Crater and P. Van Alstine, 
Phys. Rev. D1 \uline{37}, 1982 (1988)

\bibitem{cra93}
H. W. Crater, R. Becker, C. Y. Wong and P. Van Alstine, Phys. Rev.
D46, 5117 (1992).

\bibitem{tod76}
I. T. Todorov, ``Dynamics of Relativistic Point
Particles as a Problem with Constraints'', Dubna Joint
Institute for Nuclear Research No. E2-10175, 1976; Ann. Inst. H. Poincare' 
A\uline {28},207 (1978).

\bibitem{misce1}
These theta matrices have algebraic properties that permit
more efficient calculation of the
commutation relations appropriate to two spinning bodies permitting
simplification of otherwise complicated
consequences of compatibility ($[\c S_1,\c S_2]_-\psi=0$)

\bibitem{cra87}
H. W. Crater and P. Van Alstine, Phys. Rev. \uline{D36}, 3007
(1987).

\bibitem{cra90}
H. W. Crater, and P. Van Alstine, J. Math. Phys. \uline{31}, 1998  (1990).

\bibitem{saz86} 
H. Sazdjian, Phys. Rev. D1 \uline {33}, 3401(1986),
derives compatible two-body Dirac equations
but from a different starting point without the use of supersymmetry.


\bibitem{saz94}
J. Mourad and H. Sazdjian, Journal of Physics G, \uline{21}, 267 (1995).
 
\bibitem{cra94}
H. W. Crater  and P. Van Alstine, Found. Of Phys. \uline{24}, 297 (1994).

\bibitem{fnsig}
The four-vector $(\sigma_0, \bbox{\sigma})=(1,\gamma \bbox{\alpha})$
is introduced in this section  in order to write the wave function
$\Psi'$
in the simple form of Eq.\ (\ref{eq:psi}).  The quantity $\sigma_0$
here should not be confused with the $\sigma_0$ of Eq.\ (2.16).

\bibitem{barut}
A.O. Barut and S. Korny, Fortschr. Phys. \uline{33}, 309 (1985),
A.O. Barut and N. \"Unal, Fortschr. Phys. \uline{33}, 318 (1985)

\end{references}
\end{document}